\documentclass[9pt]{extarticle}
\usepackage{spconf,amsmath,graphicx,hyperref}
\usepackage{cite}
\usepackage{amsmath,amssymb,amsfonts}
\usepackage{algorithm}
\usepackage{algorithmic}
\usepackage{graphicx}
\usepackage{textcomp}
\usepackage{marvosym}
\usepackage{microtype} 
\usepackage{makecell}
\usepackage{xcolor}
\usepackage{float} 
\usepackage{booktabs} 
\usepackage{multirow} 
\usepackage{caption} 
\usepackage{graphicx} 
\usepackage{balance}
\usepackage{paralist}
\usepackage{enumitem}
\setlist[itemize]{
  noitemsep,          
  itemsep=0pt,        
  topsep=0pt,         
  leftmargin=0.5cm,   
  partopsep=0pt       
}

\title{Adaptive Speaker Embedding Self-Augmentation for Personal Voice Activity Detection with Short Enrollment Speech}
\name{Fuyuan Feng$^{1\dagger}$, Wenbin Zhang$^{2\dagger}$, Yu Gao$^{2*}$\thanks{${\dagger}$ Equal contribution. ${*}$ Corresponding author.}, Longting Xu$^{1}$, Xiaofeng Mou$^{2}$, Yi Xu$^{2}$}
\address{$^{1}$ College of Information Science and Technology, Donghua University, Shanghai, China \\
           $^{2}$ AI Research Center, Midea Group (Shanghai) Co.,Ltd., Shanghai, China \\
           \Letter\ \texttt{2232069@mail.dhu.edu.cn, \{zhangwb87, gaoyu11\}@midea.com}
           }
\begin{document}
%
\maketitle

\begin{abstract}
Personal Voice Activity Detection (PVAD) is crucial for identifying target speaker segments in the mixture, yet its performance heavily depends on the quality of speaker embeddings. A key practical limitation is the short enrollment speech—such as a wake-up word—which provides limited cues. This paper proposes a novel adaptive speaker embedding self-augmentation strategy that enhances PVAD performance by augmenting the original enrollment embeddings through additive fusion of keyframe embeddings extracted from mixed speech. Furthermore, we introduce a long-term adaptation strategy to iteratively refine embeddings during detection, mitigating speaker temporal variability. Experiments show significant gains in recall, precision, and F1-score under short enrollment conditions, matching full-length enrollment performance after five iterative updates. The source code is available at \url{https://anonymous.4open.science/r/ASE-PVAD-E5D6}.
\end{abstract}
\section{Introduction}
\label{sec:intro}
Voice Activity Detection (VAD) is a technique that identifies speech and non-speech segments at the frame level\cite{chang2018temporal}, serving as a critical front-end for tasks such as automatic speech recognition (ASR)\cite{jayasimha2021personalizing}, speaker verification (SV)\cite{kang2023svvad}, and speaker diarization\cite{11044874}. A well-trained standard VAD system detects all speech segments regardless of the speaker, enhancing downstream performance by filtering noise and reducing computational costs. 
However, in multi-speaker environments, standard VAD may trigger false positives by detecting non-target speech.
To address this limitation, Personal Voice Activity Detection (PVAD) is proposed\cite{ding2019personal}, which specifically identifies a target speaker’s voice activity in multi-talker scenarios.

PVAD systems typically operate in two stages: 1) extracting target speaker embeddings (e.g., i-vectors or d-vectors) from enrollment speech using a pre-trained speaker verification (SV) model\cite{he2021target}, and 2) utilizing these embeddings to identify target speaker segments in mixed speech~\cite{ding2019personal, jayasimha2021personalizing, kang2023svvad}. This framework simply concatenates the target speaker's embedding with the input acoustic features. Subsequent improvements introduced feature-wise linear modulation (FiLM) to adaptively condition speaker embeddings~\cite{ding2022personal, perez2018film}. While PVAD performance critically depends on embedding quality and enrollment duration, practical deployment scenarios are typically constrained to a short enrollment (e.g., wake-up words) that offer only limited speaker cues\cite{zeng2024efficient}, posing significant challenges for conventional approaches. Recent related advances have focused on leveraging enrolled audio information more effectively. This includes employing superior models (e.g., ECAPA, CAM++, WavLM) or multi-scale methods to extract higher-quality speaker embeddings from enrolled speech\cite{11011161, he2024hierarchical, zhang2025multi}, or using synthesis models to increase the duration of enrolled speech\cite{huang2025augmenting}. Some researchers have also explored the correlation between target speaker extraction (TSE) or VAD tasks and PVAD tasks, facilitating task interaction to achieve improvements\cite{sun2025towards, 11012711, 11044874, 10446581, jalal2025robust}.

\begin{figure}[t] %
    \centering
    \includegraphics[width=\columnwidth]{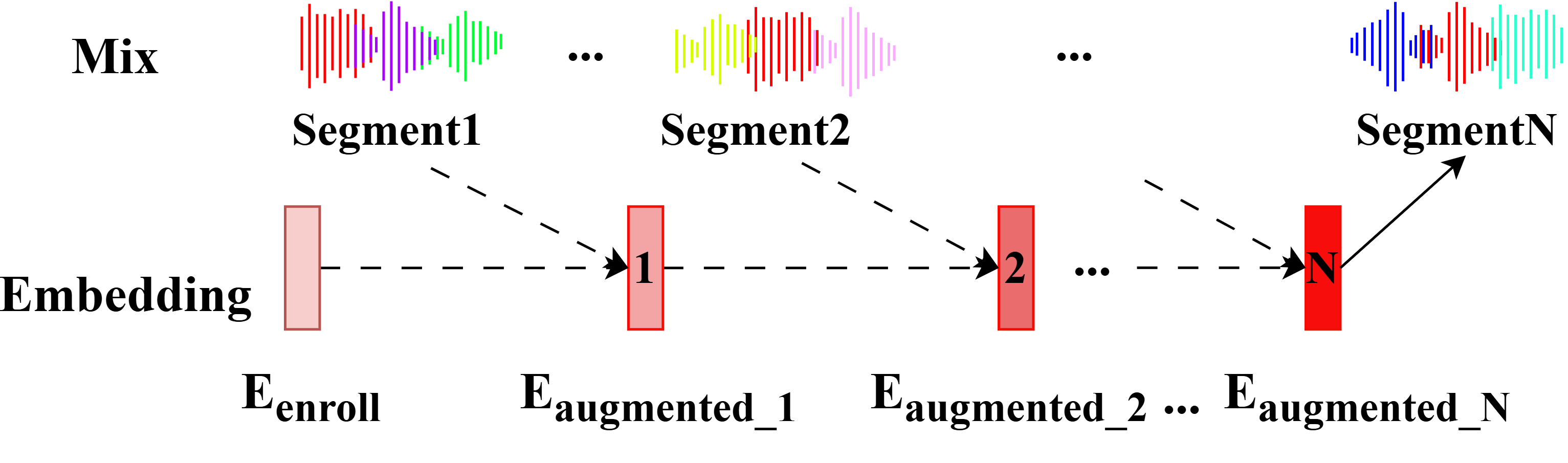} 
    \caption{Adaptive speaker embedding self-augmentation (N Iterations)} 
    \label{fig1} 
    \vspace{-2mm}
\end{figure}

\begin{figure*}[t!] 
    \centering
    \includegraphics[width=\textwidth]{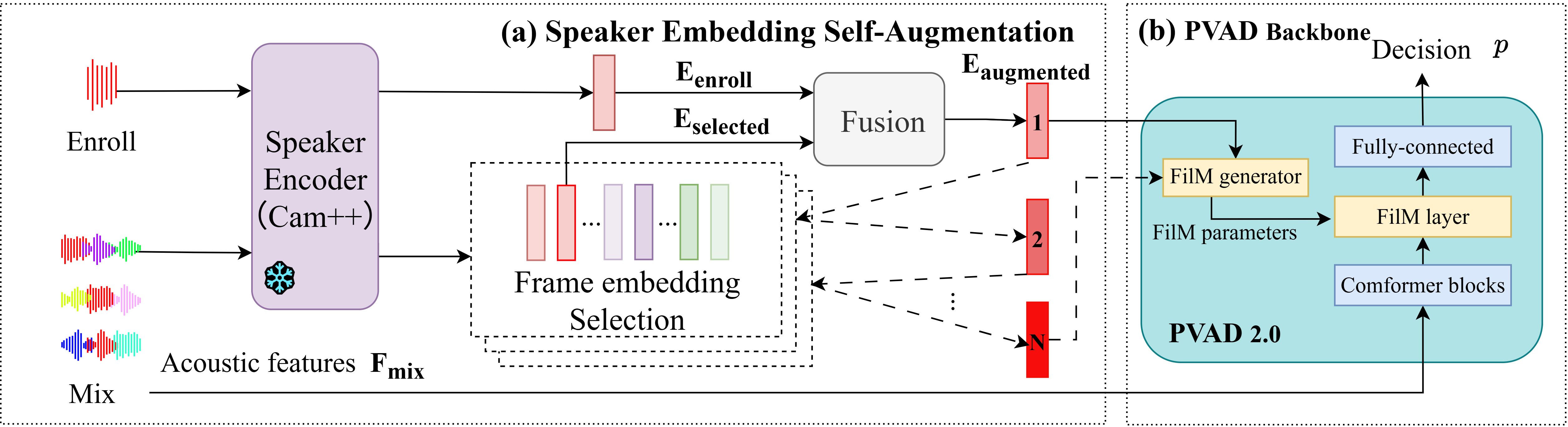} 
    \caption{Architecture of adaptive speaker embedding self-augmentation for personal VAD. (a) illustrates our self-augmentation method, where $\mathit{E}_{\mathit{augmented}}$ ((1-N) denotes the iterations of self-augmentation, as illustrates in Fig.~\ref{fig1}) is derived by similarity-guided fusion of the original enrollment embedding $\mathit{E}_{\mathit{enroll}}$ and the optimally matched frame-level embedding $\mathit{E}_{\mathit{selected}}$ from the mixed speech. (b) illustrates our PVAD 2.0 backbone architecture processing both the augmented speaker embedding \(\mathit{E}_{\mathit{augmented}}\) and the acoustic features \(\mathit{F}_{\mathit{mix}}\) extracted from the mixed speech.} 
    \label{fig2} 
 \vspace{-3mm}
\end{figure*}
A critical yet understudied consideration is that target speech inherently exists within the mixed speech input. Existing methods have employed shared encoders or cross-attention mechanisms to utilize mixed speech information, typically treating enrolled embeddings as key and value~\cite{tripathi2025attention,11012711}. However, these approaches fail to explicitly capture additional speaker identity information from the mixture. Furthermore, performance degradation occurs in scenarios like smart homes, where a domain mismatch exists between near-field enrolled audio (typically recorded on mobile devices) and far-field mixed speech. This underscores the need for explicit modeling of mixed speech characteristics. Our contributions are as follows:


\begin{itemize}
\item We propose a novel framework to augment short-enrollment embeddings, which leverages target-speaker frame embeddings extracted from mixed speech. These embeddings better retain natural acoustic properties and capture more speaker characteristics.
\item Recognizing PVAD's inherent long-term nature (as illustrated in Fig.~\ref{fig1}), we introduce iterative speaker embedding self-augmentation to dynamically accommodate speaker state variations~\cite{qin2022cross} and ensure genuine personalization.
\item Experimental results validate our iterative embedding update approach using mixed speech, demonstrating both enriched representation and the feasibility of long-term adaptive PVAD.
\end{itemize}
\vspace{-2mm}
\section{Proposed Methods}
\label{sec:Proposed Methods}
This section details proposed speaker embedding self-augmentation framework for PVAD. We first review the PVAD 2.0 architecture\cite{ding2022personal}, then comprehensively analyze the proposed speaker embedding self-augmentation component integrated into this framework.

\subsection{PVAD Backbone}
As illustrated in the PVAD Backbone section (b) of Fig.~\ref{fig2}, the model follows a standard embedding-conditioned PVAD architecture. The model takes two inputs: 1) \(\mathit{E}_{\mathit{enroll}}\): target speaker embedding extracted from enrollment audio using a pre-trained SV model, and 2) \(\mathit{F}_{\mathit{mix}}\): acoustic features derived from mixed speech containing multiple speakers.

\begin{equation}
   p = \mathit{PVAD}\bigl( \mathit{F}_{\mathit{mix}},\mathit{E}_{\mathit{enroll}} \bigr)
\end{equation}

A four-layer stacked Conformer processes the  \(\mathit{F}_{\mathit{mix}}\), while FilM modulates the \(\mathit{E}_{\mathit{enroll}}\)\cite{perez2018film}.
Finally, the fused features are fed into a fully connected layer to produce the decision $p$ of the PVAD model. The whole process can be formulated as:
\begin{equation}
    p = \mathit{FC}\Big( \mathit{FilM}\big( \mathit{Con}(\mathrm{F}_{\mathit{mix}}),\mathrm{E}_{\mathit{enroll}} \big) \Big)
\end{equation}
where $p$, \(\mathit{Con}\) denotes the posterior of the target speaker’s speech and Conformer blocks, respectively.

\begin{table*}[t!]
\centering
\renewcommand{\arraystretch}{1} 
\caption{Results for speaker embedding self-augmentation under different short enrollment speech scenarios in segment 1. Augmented\_cat/add denote concatenation/additive fusion (Eq.~\ref{eq3}/~\ref{eq4}). Metrics: accuracy(ACC), recall (REC), precision (PRE), F1-score (F1), and average precision (AP) for non-speech (NS), non-target (NTSS), and target speech (TSS).}
\vspace{-2mm}
\resizebox{\textwidth}{!}{ 
\begin{tabular}{c| c| c| c| ccccc c ccccc}
\toprule
\multirow{3}{*}{\textbf{Enroll}}& \multirow{3}{*}{\textbf{Exp}} & \multirow{1.5}{*}{\textbf{Augment}}& \multirow{1.5}{*}{\textbf{Embed}}& \multicolumn{5}{c}{\textbf{3 Sperkers ( Segment 1 )}}  & & \multicolumn{5}{c}{\textbf{3 Speakers + Noise ( Segment 1 )}} \\
\cline{5-9}\cline{11-15}
& &  \multirow{1.5}{*}{\textbf{Embed}} & \multirow{1.5}{*}{\textbf{Source}} &\multirow{2}{*}{\textbf{ACC}} &\multirow{2}{*}{\textbf{REC}} & \multirow{2}{*}{\textbf{PRE}} &\multirow{2}{*}{\textbf{F1}} & \multicolumn{1}{c}{\textbf{AP}}
 & &\multirow{2}{*}{\textbf{ACC}} & \multirow{2}{*}{\textbf{REC}} & \multirow{2}{*}{\textbf{PRE}} &\multirow{2}{*}{\textbf{F1}} & \multicolumn{1}{c}{\textbf{AP}}  \\
& & & & & & & &\textbf{[NS  NTSS  TSS]} & & & & & &\textbf{[NS  NTSS  TSS]}\\
\midrule
\multirow{4}{*}{1.5s} & E1 &$\times$ & Enroll & 86.26 & 74.85 & 86.54 & 85.22 & [\textbf{92.74} 95.40 89.41] & & 82.77 & 70.72 & 82.78 & 81.39 & [\textbf{88.46} 93.77 85.95] \\ 
                       & E2 &$\times$ & Selected & 87.14 & 79.84 & 87.38 & 86.12 & [91.91 96.05 91.05] & & 82.68 & 73.28 & 82.28 & 81.36 & [87.34 93.48 85.53] \\ 
                       & E3& $\checkmark$ & Augmented\_cat & 87.23 & 80.37 & 87.59 & 86.22 & [91.61 96.48 92.27] & & 83.45 & 76.17 & \textbf{83.36} & 82.11 & [86.40 94.59 88.70] \\ 
                       & E4& $\checkmark$ & Augmented\_add & \textbf{87.85} & \textbf{81.71} & \textbf{88.19} & \textbf{86.87} & [91.63 \textbf{96.64} \textbf{92.78}] & & \textbf{83.64} & \textbf{76.64} & 83.18 & \textbf{82.34} & [86.71 \textbf{94.73} \textbf{88.78}] \\ 

\cmidrule(lr){1-15}                     
\multirow{4}{*}{1s}   & E5 &$\times$ & Enroll & 84.74 & 71.26 & 83.96 & 83.66 & [\textbf{92.72} 94.18 86.37] & & 81.39 & 67.20 & 80.40 & 79.95 & [\textbf{88.42} 92.56 82.91] \\ 
                      & E6 &$\times$ &  Selected & 86.52 & 78.51 & 86.28 & 85.51 & [91.91 95.43 89.50] & & 82.38 & 72.68 & 81.70 & 81.06 & [87.33 93.08 84.81] \\ 
                      & E7 & $\checkmark$ & Augmented\_cat & 86.28 & 77.98 & 86.35 & 85.25 & [91.58 95.70 90.50] & & 82.68 & 73.94 & 82.36 & 81.30 & [86.39 93.96 87.02] \\ 
                      & E8 & $\checkmark$ & Augmented\_add  & \textbf{87.14} & \textbf{80.09} & \textbf{87.15} & \textbf{86.16} & [91.61 \textbf{96.19} \textbf{91.73}] & & \textbf{83.22} & \textbf{74.50} & \textbf{83.20} & \textbf{81.75} & [86.81 \textbf{94.27} \textbf{87.67}] \\

\cmidrule(lr){1-15}                    
\multirow{4}{*}{0.5s} & E9 &$\times$ & Enroll & 81.56 & 63.00 & 78.90 & 80.28 & [\textbf{92.46} 91.30 78.98] & & 78.42 & 58.94 & 75.80 & 76.70 & [\textbf{88.23} 89.47 75.43] \\ 
                      & E10 &$\times$&  Selected & 84.88 & \textbf{75.78} & 82.67 & 83.92 & [91.96 93.56 85.08] & & 80.21 & \textbf{68.44} & 77.35 & 78.92 & [87.37 90.67 78.73] \\ 
                      & E11 & $\checkmark$ & Augmented\_cat & 84.03 & 72.89 & 82.68 & 82.94 & [91.40 94.22 86.55] & & 80.08 & 67.61 & 78.39 & 78.62 & [86.09 91.78 81.77] \\ 
                      & E12 & $\checkmark$ & Augmented\_add  & \textbf{85.34} & 75.34 & \textbf{84.68} & \textbf{84.3} & [91.54 \textbf{94.83} \textbf{88.39}] & & \textbf{80.79} & 68.12 & \textbf{80.05} & \textbf{79.31} & [86.80 \textbf{92.47} \textbf{83.38}] \\ 
\bottomrule
\end{tabular}
}
\label{tab1}
\vspace{-2mm}
\end{table*}

\subsection{Speaker Embedding Self-Augmentation}
\label{ssec:Self-Augmentation}
\subsubsection{Self-Augmentation}

As illustrated in (a) of Fig.~\ref{fig2}, we first perform a long-window frames level speaker embedding similarity calculation on the mixed speech, similar to a low temporal resolution PVAD detection. Mixed speech is segmented with a frame length of 1s and a frame shift of 0.2s, and the speaker encoder is performed using a pre-trained model \(\text{CAM++}\)\textsuperscript{\footnotemark}\footnotetext{\url{https://github.com/modelscope/3D-Speaker}} to obtain \(\mathit{E}_{\mathit{mix}} \in \mathbb{R}^{T \times D}\) and \(\mathit{E}_{\mathit{enroll}} \in \mathbb{R}^{1 \times D}\) \cite{wang2023cam++, chen20253d}. Then calculate the cosine similarity between \(\mathit{E}_{\mathit{mix}} \in \mathbb{R}^{T \times D}\) and \(\mathit{E}_{\mathit{enroll}} \in \mathbb{R}^{1 \times D}\) along the T dimension, and find the key frame’s embedding that exceeds the set threshold and has the highest similarity, which we named \(\mathit{E}_{\mathit{selected}} \in \mathbb{R}^{1 \times D}\). 

Although overlapping speakers may have an impact, in real conversation data, the speech of a single speaker usually occupies most of the time, and we only use one key frame in the mixed speech, greatly reducing the probability of being affected. Therefore, we have reason to believe that there are target speaker information in \(\mathit{E}_{\mathit{selected}}\) that is not completely similar to \(\mathit{E}_{\mathit{enroll}}\), and is more realistic and more representative of the actual mixed acoustic environment.

Then, we perform feature fusion on \(\mathit{E}_{\mathit{selected}}\) and \(\mathit{E}_{\mathit{enroll}}\). Based on the research of \cite{tripathi2025attention}, we compared the simpler methods of addition and concatenation as fusion methods. The fused features are named \(\mathit{E}_{\mathit{augmented}}\):

\begin{equation}
   \mathit{E}_{\mathit{augmented\_cat}} = [\mathit{E}_{\mathit{enroll}}; \mathit{E}_{\mathit{selected}}]
    \label{eq3}
\end{equation}

\begin{equation}
   \mathit{E}_{\mathit{augmented\_add}} = \mathit{E}_{\mathit{enroll}} + \mathit{E}_{\mathit{selected}}
   \label{eq4}
\end{equation}
This is the process of speaker embedding self-augmentation.

\subsubsection{Long-term Speaker Embedding Adaptation}
\label{ssec:Update}
The proposed speaker embedding self-augmentation from mixed speech addresses short-term needs, but activity detection is a long-term process (as illustrated in Fig.~\ref{fig1}).  A static enrollment audio may inadequate for detection as it fails to capture the speaker's temporal vocal dynamics. We validate a long-term speaker embedding update strategy, where the reference embedding evolves progressively during PVAD. Specifically: 


    
    
    


\begin{itemize}
\item After processing Segment 1, the reference updates from \(\mathit{E}_{\mathit{enroll}}\) to \(\mathit{E}_{\mathit{augmented\_1}}\).
\item For Segment 2, we compute similarity between its Frame-level embedding and \(\mathit{E}_{\mathit{augmented\_1}}\), then fuse the selected \(\mathit{E}_{\mathit{selected\_2}}\) with \(\mathit{E}_{\mathit{augmented\_1}}\) to yield \(\mathit{E}_{\mathit{augmented\_2}}\).
\item This iterative process continues until Segment N, transforming the initial \(\mathit{E}_{\mathit{enroll}}\) into \(\mathit{E}_{\mathit{augmented\_N}}\).
\end{itemize}

In order to achieve more robust embeddings in long-term fusion, we reformulate the long-term speaker embedding fusion and update rule as follows:
\begin{equation}
   \mathit{E}_{\mathit{avg\_add}} = 0.5×(\mathit{E}_{\mathit{augmented\_n-1}} + \mathit{E}_{\mathit{selected}})
   \label{eq5}
\end{equation}
\begin{equation}
   \mathit{E}_{\mathit{augmented\_n}} = \mathrm{\lambda}×\mathit{E}_{\mathit{enroll}} +(1-\mathrm{\lambda})×\mathit{E}_{\mathit{avg\_add}}
   \label{eq6}
\end{equation}
when n=1, \(\mathit{E}_{\mathit{enroll}}\) is used for \(\mathit{E}_{\mathit{augmented\_n-1}}\). In this way, we obtain more enrolled speaker information in the long term detection, which are at different time points, and change the reference information from a brief \(\mathit{E}_{\mathit{enroll}}\)  to a rich and accurate \(\mathit{E}_{\mathit{augmented\_N}}\).

\section{Experimental Setup}
\label{sec:Experiments}

\subsection{Datasets}
\label{ssec:Datasets}

Following the experimental setup in~\cite{ding2019personal,xu2024personal} for conversational speech simulation, we construct a multi-speaker training set using LibriSpeech's 500-hour subsets~\cite{panayotov2015librispeech}, comprising 1,166 speakers. We generate 99,800 concatenated utterances, each containing three speakers with one randomly selected as the target.  Frame-level PVAD labels were obtained using the Montreal Forced Aligner~\cite{mcauliffe2017montreal}, where speech segments were classified as either target or non-target speaker activity.

Additionally, we create ten evaluation sets (simulate 10 mixed segments (N=10 in Fig.~\ref{fig1})) from LibriSpeech's test-clean and test-other subsets, each containing approximately 3,700 concatenated utterances matching the training configuration. We augment both the train and test sets by adding noise from the MUSAN corpus\cite{snyder2015musan} at randomized signal-to-noise ratios (SNR) between 0-20 dB to simulate realistic acoustic conditions. 

\subsection{Implementation Details}
We adopt PVAD2.0~\cite{ding2022personal} as baseline, where exclusively the 192-dim \(\mathit{E}_{\mathit{enroll}}\) extracted from the enrolled audio (via pre-train CAM++) serves as a reference. In all models, we input acoustic features \(\mathit{F}_{\mathit{mix}}\) with the same settings, i.e., 40-dim log Mel-filterbank energy with a frame length of 25 ms and a frame shift of 10 ms. The loss function uses Weighted Pairwise Loss\cite{ding2019personal}, where the weights are set to w⟨tss, ns⟩ = w⟨tss, ntss⟩ =1 and w⟨ns, ntss⟩ = 0.5. During the training process, we use the Adam optimizer\cite{kingma2014adam} with an initial learning rate of 5e-4. The weighting parameter $\lambda$ in Eq.~\ref{eq6} was set to 0.1. We trained for 100 epochs and selected the best performing checkpoint as the final result.

The \(\mathit{E}_{\mathit{enroll}}\) during training is extracted from the enrolled audio at 1.5s, while the \(\mathit{E}_{\mathit{enroll}}\) during testing is extracted from the enrolled audio at 0.5s, 1s, and 1.5s, ensuring that they correspond to the same audio.
We evaluate performance using standard metrics: accuracy (ACC), recall (REC), precision (PRE), F1-score (F1), and class-wise average precision (AP) for non-speech (NS), non-target speaker speech (NTSS), and target speaker speech (TSS).

\begin{table*}[t!]
\centering
\renewcommand{\arraystretch}{1} 
\caption{Results for speaker embedding self-augmentation under different short enrollment speech scenarios in segment 2. 
Augmented\_1 is the embedding augmented once using Segment 1 and
Augmented\_2 is Augmented\_1 further augmented using Segment 2, with both methods using additive fusion(Eq.~\ref{eq4}).}
\vspace{-2mm}
\resizebox{\textwidth}{!}{ 
\begin{tabular}{c| c| c| c| ccccc c ccccc}
\toprule
\multirow{3}{*}{\textbf{Enroll}} &\multirow{3}{*}{\textbf{Exp}} & \multirow{1.5}{*}{\textbf{Augmnet}}& \multirow{1.5}{*}{\textbf{Embed}}& \multicolumn{5}{c}{\textbf{3 Sperkers ( Segment 2 )}}  & & \multicolumn{5}{c}{\textbf{3 Speakers + Noise ( Segment 2 )}} \\
\cline{5-9}\cline{11-15}
& &\multirow{1.5}{*}{\textbf{Embed}}& \multirow{1.5}{*}{\textbf{Source}} &\multirow{2}{*}{\textbf{ACC}} &\multirow{2}{*}{\textbf{REC}} & \multirow{2}{*}{\textbf{PRE}} &\multirow{2}{*}{\textbf{F1}} & \multicolumn{1}{c}{\textbf{AP}}
 & &\multirow{2}{*}{\textbf{ACC}} & \multirow{2}{*}{\textbf{REC}} & \multirow{2}{*}{\textbf{PRE}} &\multirow{2}{*}{\textbf{F1}} & \multicolumn{1}{c}{\textbf{AP}}  \\
&& & & & & & &\textbf{[NS  NTSS  TSS]} & & & & & &\textbf{[NS  NTSS  TSS]}\\
\midrule
\multirow{3}{*}{1.5s} & E13 &$\times$& Enroll  &  87.00 & 75.99 & 86.79 & 86.10 & [\textbf{95.00} 95.99 90.23]&& 82.18 & 69.58 & \textbf{81.35} & 80.77 & [88.29 93.31 84.46]\\ 
                      & E14 & $\checkmark$&  Augmented\_1    & 87.47 & 79.59 & \textbf{87.27} & 86.57 & [93.59 96.57 92.00]&& 82.70 & 73.07 & 81.22 & 81.40 & [\textbf{88.48} 94.22 86.61]\\ 
                      & E15 &$\checkmark$&  Augmented\_2 &  \textbf{88.93} & \textbf{88.32}& 84.59 & \textbf{88.29} & [93.33 \textbf{97.35} \textbf{93.85}]&& \textbf{83.94} & \textbf{83.99} & 77.24 & \textbf{82.98} & [88.15 \textbf{95.11} \textbf{89.04}]\\ 
                      
\cmidrule(lr){1-15}                     
\multirow{3}{*}{1s}   & E16 &$\times$&  Enroll    & 85.47 & 72.33 & 84.19 & 84.52 & [\textbf{94.90}  94.71 86.85]&& 80.55 & 65.39 & 78.68 & 79.04 & [88.10  91.94 80.96]\\ 
                      & E17 &$\checkmark$& Augmented\_1   &  85.95 & 76.09 & \textbf{84.78} & 85.03 & [93.64 95.46 89.25]&& 81.89 & 69.84 & \textbf{80.87} & 80.50 & [\textbf{88.59} 93.50 85.04]\\ 
                      & E18 &$\checkmark$& Augmented\_2   &  \textbf{87.82} & \textbf{86.12} & 82.82 & \textbf{87.23} & [93.39 \textbf{96.56} \textbf{91.98}]&& \textbf{83.55} & \textbf{81.98} & 77.34 & \textbf{82.56} & [88.31 \textbf{94.74} \textbf{88.06}]\\

\cmidrule(lr){1-15}                    
\multirow{3}{*}{0.5s} & E19 &$\times$&  Enroll   &  82.89 & 64.90 & 80.82 & 81.71 & [\textbf{94.76} 92.36 81.01]&& 78.08 & 58.70 & 74.65 & 76.35 & [88.01 89.24 74.37]\\ 
                      & E20 &$\checkmark$& Augmented\_1  &  85.40 & 73.25 & \textbf{85.34} & 84.35 & [93.61  95.20 88.76]&& 80.43 & 65.61 & \textbf{78.96} & 78.90 & [\textbf{88.49}  92.05 81.21]\\
                      & E21 &$\checkmark$& Augmented\_2  &  \textbf{87.55} & \textbf{84.91} & 82.88 & \textbf{86.93} & [93.34 \textbf{96.45} \textbf{91.77}]&& \textbf{82.25} & \textbf{78.85} & 75.51 & \textbf{81.32} & [88.37 \textbf{93.61} \textbf{85.22}]\\ 
\bottomrule
\end{tabular}
}
\label{tab2}
\vspace{-3mm}
\end{table*}

\begin{figure*}[t!] 
    \centering
\includegraphics[width=\textwidth]{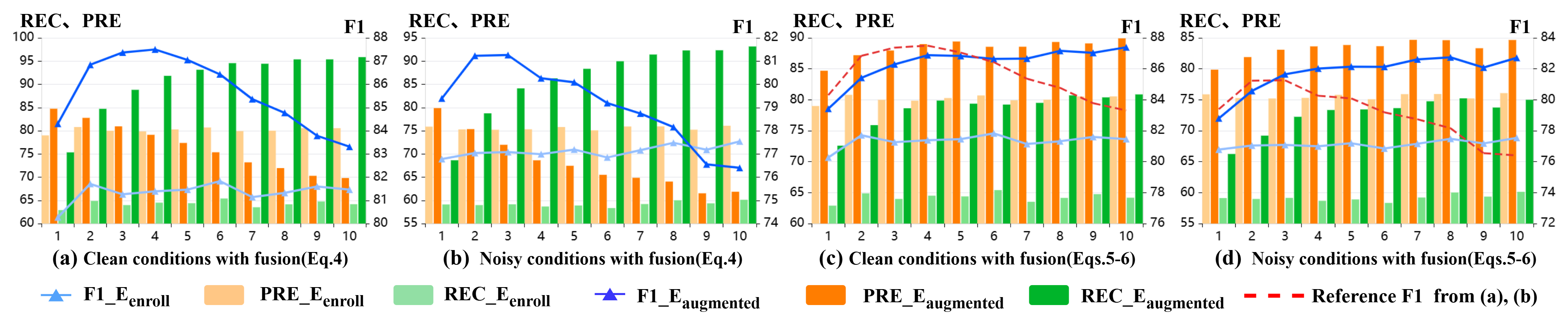} 
    \caption{Results for speaker embedding self-augmentation with 0.5s enrollment in segment 1-10 (where N=10, as defined in Fig.~\ref{fig1}). Bars (left axis) show REC/PRE; lines (right axis) show F1; X-axis shows Segment 1-10. Experimental conditions: (a) Clean, Eq.~\ref{eq4} fusion; (b) Noisy, Eq.~\ref{eq4} fusion; (c) Clean, Eqs.~\ref{eq5}--\ref{eq6} fusion; (d) Noisy, Eqs.~\ref{eq5}--\ref{eq6} fusion. For comparison, the red dashed line in (c) and (d) replicates the F1 curve from (a) and (b), respectively.
} 
    \label{fig3} 
    \vspace{-3mm}
\end{figure*}

\section{Results and Discussion}
\label{sec:Results}
\subsection{Speaker Embedding Self-Augmentation}
\subsubsection{Performance in Segment 1}

Table~\ref{tab1} compares PVAD performance under three short enrollment for mixed speech Segment 1. The results (E1vsE2, E5vsE6, E9vsE10) demonstrate that the keyframe (1s) embedding \(\mathit{E}_{\mathit{selected}}\) selected from mixed speech achieves competitive performance across most metrics, even surpassing 1.5s enrollment audio in some cases, as it better matches the acoustic environment of mixed speech. Regarding the two fusion approaches (Eq.~\ref{eq3} and Eq.~\ref{eq4}), our experimental results align with \cite{tripathi2025attention} in demonstrating that the additive fusion method achieves an optimal trade-off, exhibiting only marginal degradation in NS accuracy (-1.32 average) while delivering substantial improvements in both NTSS (+2.08 average) and TSS (+5.61 average) performance. These findings validate its effectiveness, and consequently, we adopt the additive fusion as the default approach subsequent experiments.

\vspace{-2mm}
\subsubsection{Performance in Segment 2}
\label{Performance in segment 2}
Table~\ref{tab2} evaluates three short enrollment scenarios for Segment 2. Two key findings emerge:
\begin{itemize}
\item \textbf{Embedding quality enhancement:} The performance of \(\mathit{E}_{\mathit{augmented\_1}}\) (E14/E17/E20) consistently surpass original \(\mathit{E}_{\mathit{enroll}}\), in agreement with the results presented in Table~\ref{tab1}, further demonstrating the effectiveness of our self-augmentation approach.
\item \textbf{Iterative gain:} Further improvement in recall (+15.4\% avg) and F1-score (+2.09 avg) is achieved through secondary augmentation (\(\mathit{E}_{\mathit{augmented\_2}}\), E15/E18/E21), demonstrating cumulative benefits from progressive refinement.
\end{itemize}

These experimental findings demonstrate the feasibility of long-term self-augmentation in speaker embedding systems.


\begin{figure}[b!] 
\vspace{-4mm}
    \centering
\includegraphics[width=\columnwidth]{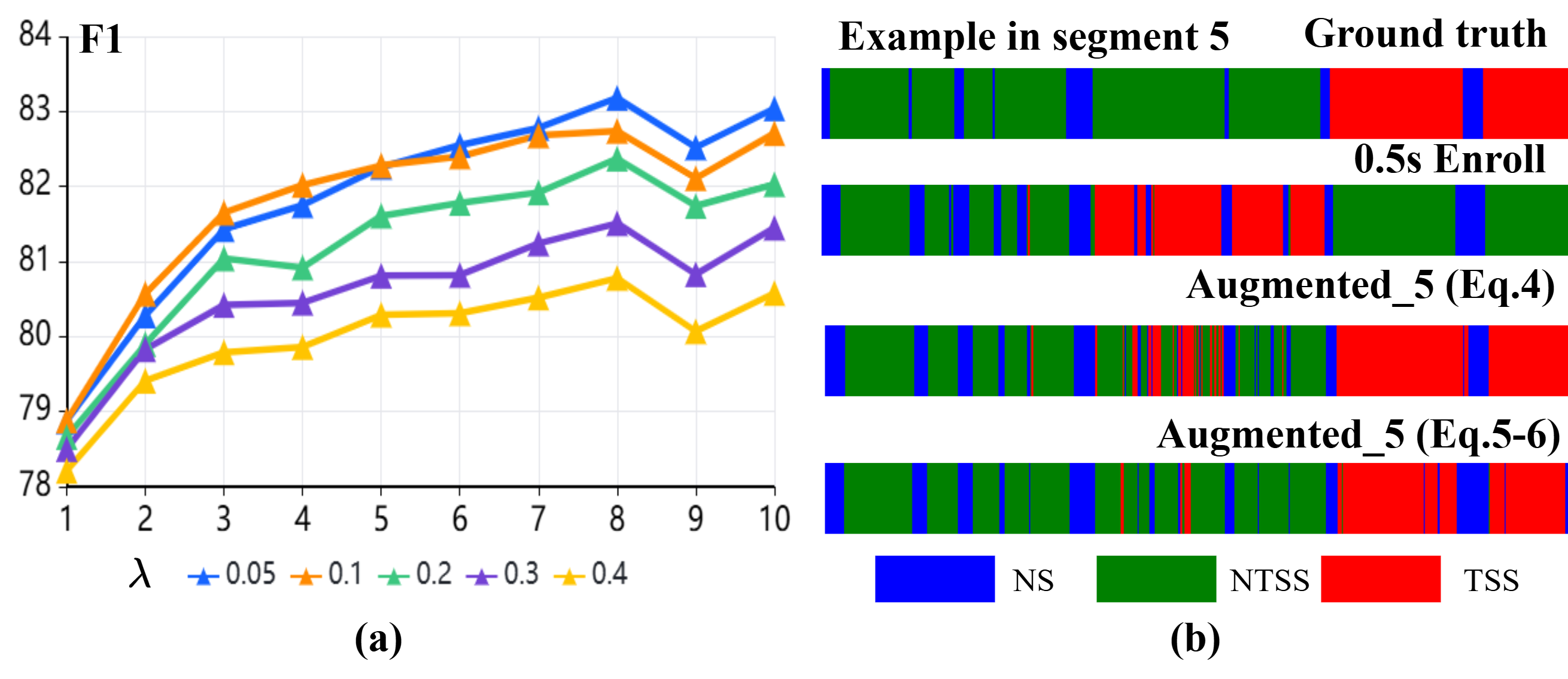} 
    \caption{(a) Ablation study of $\lambda$ (0.05, 0.1, 0.2, 0.3, 0.4) in Eq.~\ref{eq6} for speaker embedding self-augmentation under 0.5s enrollment over 1-10 iterations (where N=10 in Fig.~\ref{fig1}) in noisy condition. (b) PVAD performance visualization comparison: Original 0.5s enrollment vs. 5-iteration augmented embeddings (Eq.~\ref{eq4} vs. Eqs.~\ref{eq5}--\ref{eq6}).} 
    \label{fig4} 
    \vspace{-2mm}
\end{figure}
\vspace{-1mm}
\subsection{Long-term Speaker Embedding Adaptation}
\subsubsection{Performance in Segment 1-10}
Fig.~\ref{fig3} compares progressive enhancement of 0.5s enrollment speech across ten segments in clean (a,c) versus noisy (b,d) environments, using Eq.~\ref{eq4} for (a,b) and Eqs.~\ref{eq5}--\ref{eq6} for (c,d). The fusion strategy defined by Eq.~\ref{eq4} , as visualized in subfigures (a) and (b) of Fig.~\ref{fig3}, results in a significant decrease in both precision and F1-score with more iterations of augmentation. We attribute this to the introduction of noise (which, in the clean scenario, refers to the speech of non-target speakers) in the selection of \(\mathit{E}_{\mathit{selected}}\), which is then amplified through iterative error accumulation.

As shown in (c) and (d), the modified fusion strategy (Eqs.~\ref{eq5}--\ref{eq6}), incorporating weighted averaging and residual correction with enrolled embeddings, demonstrates consistent improvements in precision, recall, and F1-score across both conditions. Remarkably, \(\mathit{E}_{\mathit{augmented\_5}}\) achieves performance parity with the embedding extracted from the full enrollment speech (7.4s avg) after 5 iterations, with this gain attributable exclusively to embedding-level augmentation (model parameters remain fixed throughout the process).

\subsection{Ablation Study of $\lambda$ (Eq.~\ref{eq6}) and Visualization}


Fig.~\ref{fig4} (a) presents the F1-scores over 1–10 long-term iterations using Eqs.~\ref{eq5}--\ref{eq6} with various values of $\lambda$. Consistent with earlier findings, iterations of speaker embedding enhancement typically improve the F1-scores. $\lambda \in [0.05, 0.1]$ yields the best performance.

Fig.~\ref{fig4} (b) compares PVAD performance on a Segment 5 sample using embeddings from: the original 0.5s enrollment, 5-iteration enhancement with Eq.~\ref{eq4}, and 5-iteration enhancement using Eqs.~\ref{eq5}--\ref{eq6}. The short enrollment yields poor performance due to limited information, causing same-gender speaker confusion and detection errors. Iterative embedding enrichment mitigates this by providing more discriminative features, substantially improving same-gender differentiation and overall PVAD performance, especially with Eqs.~\ref{eq5}--\ref{eq6}.

\vspace{-3mm}
\section{Conlusion}
\label{sec:Conlusion}

We propose a speaker embedding self-augmentation framework to address enrollment data scarcity in PVAD systems. By extracting and fusing target embeddings from key frames in mixed speech, our method enhances enrollment robustness while adapting to dynamic acoustics. An iterative update strategy ensures stable long-term detection without extra parameters. Experiments show significant gains in recall, precision, and F1-score, especially for short enrollments, demonstrating embedding-level augmentation as an effective solution for personalized speech processing with limited data.

\clearpage

\bibliographystyle{IEEEbib}
\bibliography{refs}

\end{document}